\documentstyle[prb,aps,preprint]{revtex}

\begin{document}
\draft
\title{Ghost spins and novel quantum critical behavior in a spin chain with local bond-deformation}
\author{Jianhui Dai$^{a,b}${~~~} Yupeng Wang$^{c,d}${~~~} U. Eckern$^c$}
\address{
$^a$Abdus Salam International Center for Theoretical Physics, Trieste 34100, Italy\\
$^b$Zhejiang Institute of Modern Physics, Zhejiang University, Hangzhou 310027, China\\
$^c$Institut f\"{u}r Physik, Universit\"{a}t Augsburg, D-86135, Augsburg, Germany\\
$^d$Cryogenic Laboratory, Chinese Academy of Sciences, Beijing 100080,  China}

\maketitle
\begin{abstract}
  We study the boundary impurity-induced critical behavior in an integrable $SU(2)$-invariant model consisting 
  of an open Heisenberg chain of arbitrary spin-$S$ (Takhatajian-Babujian model) interacting with an impurity 
  of  spin $\vec{S'}$ located at one of the
 boundaries.  For $S=1/2$ or $S'=1/2$, the impurity interaction has a very simple form $J\vec{S}_1\cdot\vec{S'}$ 
 which describes the deformed boundary bond between the impurity $\vec{S'}$ and the first bulk spin $\vec{S}_1$ with
  an arbitrary strength $J$. With a weak coupling $0<J<J_0/[(S+S')^2-1/4]$, the impurity is completely compensated, 
  undercompensated, and overcompensated for $S=S'$, $S>S'$ and $S<S'$ as in the usual Kondo problem.  While for strong 
  coupling $J\geq J_0/[(S+S')^2-1/4]$, the impurity spin is split into two ghost spins. Their cooperative effect leads 
  to a variety of new critical behaviors with different values of $|S'-S|$.
\end{abstract}


\section{Introduction}
 Quantum fluctuations induced by an impurity coupled to the one-dimensional Tomonaga-Luttinger 
 liquid (TLL) play essential role in understanding the low temperature behavior of quasi-1D systems, 
 such as quantum wires\cite{1}, fractional quantum Hall effect\cite{2}, 
 carbon nanotubes \cite{3} or quasi-1D organic conductors\cite{4}. The problem 
of an impurity spin $\vec{S'}$ coupled  with both of the neighboring sites of quantum chains
 was studied by a class of integrable $SU(2)$-invariant models\cite{5,6}. For the 
Heisenberg chain with ferromagnetic coupling, the impurity is locked into the critical behavior of 
the lattice, i.e., at low temperatures the specific heat is proportional to $T^{1/2}$ and the 
susceptibility diverges as $T^{-2}$ with logarithmic corrections\cite{7}. For a chain with 
antiferromagnetic coupling, the impurity spin is compensated by bulk spins with three different 
situations similar to the multichannel Kondo problem\cite{8}: for $S'=S$, it is the complete 
compensation and the impurity just corresponds to one more site in the chain; for $S'>S$, 
the partial compensation with Schottky anomaly when an external magnetic field $H$ is applied;
and for $S'<S$, the overcompensation which gives rise to quantum critical behavior\cite{8}.

 Meanwhile the effects of an impurity embedded in 1D TLL 
 were extensively discussed recently. By renormalization group(RG) techniques, bosonization methods 
 and boundary conformal field theories, many interesting results 
have been obtained, showing  the unusual properties of TLL in the presence of a local potential barrier or
a magnetic impurity\cite{9,10,11,12}. Generally speaking, these new findings indicate that the quantum 
impurity models renormalize to critical points
 corresponding to conformally invariant boundary conditions\cite{13,14,15,16}. The impurity-bulk coupling 
 strength $J$ flows either to infinity when the impurity is screened, or to finite as if it is overscreened,
no matter what the sign of $J$ is initially. In particular, numerical studies of the finite size spectrum 
support the picture that the fixed point corresponds to a chain disconnected at the impurity site for 
repulsive interaction\cite{17}.  
However , the low temperature impurity behavior described by previous Bethe ansatz integrable models 
do not correspond to the stable critical points mentioned above.  For instance, at the critical point 
the spin-$S$ impurity coupled to spin-$1/2$ 
antiferromagnetic Heisenberg chain has the effective screened spin $S_{eff}=S-1/2 $ rather than $S-1$,
despite the fact that it couples to the two neighboring $1/2$-spin's\cite{5}. In this 
respect, the critical point described by 
the impurity model is unstable, owing to the fact that these models have a fixed impurity coupling,
a ``fine tuned" impurity interaction term, and no backward scatterings\cite{18}. It is recalled that 
 backward scattering is one of the essence of quantum impurity problem in  1D TLL\cite{9,11}. From the 
 point of view of RG, electrons or spin waves moving in one
-dimensional space will be largely scattered back by the impurity, while the tunneling effect could 
be perturbed\cite{11,19}. In the fixed point limit, they are completely scattered back after a phase shift due to 
interaction with the impurity, as long as transmission being plausibly 
neglected at sufficiently low temperature. Based on the fixed point observed by RG, some open boundary problems
with the impurity located at the boundary were considered\cite{20,21}. By comparing the open boundary models
and the periodic models, Zvyagin and co-author\cite{20,21} argued that the impurities in the integrable
models could be generic. However, we remark either a transparent impurity or a boundary impurity in
the integrable models considered earlier can not reveal a full picture of a generic impurity in the bulk
since only one channel host is included in these models (for the transparent impurity, only forward
scattering while for the boundary impurity, only one half-chain). To make the problem to be generic,
one should consider two half-chains interacting symmetrically with the impurity\cite{22}. At low energy 
scales, the problem is effectively a two-channel one\cite{18,23}, as long as the interaction in the bulk is
repulsive. Generally, tunneling through the impurity may exist. This causes hybridization,
splitting and anisotropy of the two channels. However, the tunneling matrix is negligibly small comparing
to the Kondo coupling\cite{11} or the impurity potential\cite{9} at low energy scales from the RG point
of view. Therefore, these effects are not very harmful to the two-channel behavior\cite{24}.
\par
In this paper, we consider the problem of an impurity spin interacting symmetrically with two half spin-chains.
By mapping the problem to an open boundary one, we solve the related integrable model via Algebraic
Bethe ansatz (ABA). The structure of the present paper is the following: In the subsequent section, we
construct the model and derive the Bethe ansatz equation. The ground state properties and the boundary bound states
 will be discussed in sect. III.
In sect.IV, we discuss the thermodynamics of the open boundary as well as the impurity. It is found that the open boundary
behaves as an overscreened spin and the impurity itself, may show different quantum critical behaviors, depending on
the coupling constant $J$. Sect.V is attributed to the concluding
remarks.
\section{The model and its Bethe ansatz}
Let us start with the following Hamiltonian\cite{25}
\begin{eqnarray}
H=J_0\sum_{n=1}^{N-1}[\vec{S}_n\cdot\vec{S}_{n+1}+\vec{S}_{-n}\cdot\vec{S}_{-n-1}]+J\vec{S'}\cdot
(\vec{S}_1+\vec{S}_{-1}),
\end{eqnarray}
where $J_0>0$ (antiferromagnetic coupling); $\vec{S}_n$ is the spin-1/2 operator, $\vec{S'}$ is the impurity
spin operator. As long as $J_0\neq J$ or $S'\neq 1/2$, back scattering off the impurity dominates over
the forward scattering and the problem is effectively a two-channel one. No matter $J>0$ (antiferromagnetic)
or $J<0$ (ferromagnetic), the second term in (1) causes ferromagnetic correlation between $\vec{S}_1$
and $\vec{S}_{-1}$. That means $\vec{S}_1$ and $\vec{S}_{-1}$ favors to form a spin-1 composite at low 
temperatures. In fact, there is a quasi-long-distance ferromagnetic correlation between $\vec{S}_n$ and 
$\vec{S}_{-n}$
\begin{eqnarray}
<\vec{S}_{n}\cdot\vec{S}_{-n}>\sim (2n)^{-\theta},
\end{eqnarray}
where the exponents $\theta$ varies from 1 to 4, which can be derived from the boundary conformal filed theory\cite{19}.
Therefore, there is a tendency for $\vec{S}_{n}$ and $\vec{S}_{-n}$ to form a spin-1 composite and the 
problem could be mapped to a spin-1 chain coupled with a boundary impurity. We remark the composites
far away from the impurity are not very tight and may lose sense as can be seen in (2), but this does not change very much the impurity behavior since the bulk-impurity interaction is local and very short-ranged.
Hence, a spin-$S$ chain with a boundary impurity keeps the main feature of an impurity in a spin-$S/2$
chain. Caution should be taken when we construct the spin-$S$ open chain to avoid the Haldane gap since the original
Hamiltonian (1) is gapless. The formation of the high-spin composites is very similar to that in a multi-channel Kondo problem\cite{8} but with a different mechanism. Of course, the boundary impurity-induced critical behavior is quite generic.

Based on the above discussion, we study the low temperature behavior induced by an magnetic impurity 
$\vec{S'}$ coupled to an open antiferromagnetic Heisenberg chain of spin $S$ by use of Bethe ansatz 
exact solutions. It is well known that integrable generalization of isotropic $S=1/2$ spin chain 
to arbitrary spin $S$ leads to the Hamiltonian\cite{26}
\begin{eqnarray}
H_S=J_0\sum^{N-1}_{j=1}Q_{2S}(\vec{S}_j\cdot\vec{S}_{j+1}),
\end{eqnarray}
where $Q_{2S}(x)$ is a polynomial of degree $2S$ of $SU(2)$ invariant quantities
$x=\vec{S}_j\cdot\vec{S}_{j+1}$
\begin{eqnarray}
Q_{2S}(x)=\displaystyle{\sum^{2S}_{j=1}(\sum^{j}_{k=1}\frac{1}{k})
\prod^{2S}_{l\neq j, l=0}\frac{x-x_l}{x_j-x_l}},
\end{eqnarray} 
with $x_n=\frac{1}{2}n(n+1)-S(S+1)$, $n=0,1,\cdots, 2S$. One recovers 
$H_{1/2}=J_0\sum^{N-1}_{j=1}\vec{S}_j\cdot\vec{S}_{j+1}$, 
$H_{1}=J_0\sum^{N-1}_{j=1}[\vec{S}_j\cdot\vec{S}_{j+1}-
(\vec{S}_j\cdot\vec{S}_{j+1})^2]$
as the usual Spin-1/2 Heisenberg model and the $S=1$
Takhatajian-Babujian model respectively (up to an irrelevant constant). 
 The construction of the model is based on the vertex weight operators $R(\lambda)$, represented by 
 matrices acting on the tensor product spaces  $V_1\otimes V_2$ of two spins $\vec{S}_1$, $\vec{S}_2$,
  with a parameter $\lambda$ identified as spin rapidity.  Its explicit form is
\begin{eqnarray}
_SR^{12}(\lambda)=-\sum^{2S}_{l=0}\prod^l_{k=0}\frac{\lambda-k}{\lambda+k}
P_l,
\end{eqnarray}
where $P_l$ is the projector selecting the states with total spin $l$ in  the tensor product of the two spins involved, 
$P_l(x)=\prod^{2S}_{n\neq l,n=0}\frac{x-x_n}{x_l-x_n}$.
 Owing to the Yang-Baxter equations satisfied by the $R$-matrix, we have the relationship
$H_S\propto \frac{d}{d\lambda}\ln t(\lambda)|_{\lambda=0}$,
with $t(\lambda)$ being the transfer-matrix defined by
\begin{eqnarray}
t(\lambda)=tr_AT(\lambda)=tr_{A}\{ _SR^{AN}(\lambda) \cdots _SR^{A1}(\lambda)\}.
\end{eqnarray}  
Here, the trace is taken in the auxiliary spin space $V_A$ ($dim V_A=2S+1$)
introduced to help us track of the proliferating spin indices. Because 
$[t(\lambda), t(\mu)]=0, ~~\forall \lambda, \mu$,
$H_S$ is integrable under the periodic boundary condition. 

   Now, we put a magnetic impurity $\vec{S'}$ at one of the end of an 
{\it open} spin-S Heisenberg chain, by considering the following {\it integrable} Hamiltonian
\begin{eqnarray}
H=H_S+H_{imp},\\
H_{imp}=J_0\sum^{S+S'}_{l=|S-S'|+1}(\sum^l_{k=|S-S'|+1}\frac{k}{k^2-c^2})
\prod^{S+S'}_{n\neq l, n=|S-S'|}\frac{y-y_n}{y_l-y_n},
\end{eqnarray}
where $y=\vec{S}_1\cdot\vec{S'}$, $y_l=\frac{1}{2}[l(l+1)-S(S+1)-S'(S'+1)]$; $c$ is an arbitrary 
parameter describing the strength of the bulk-impurity interaction.
The impurity is assumed to be sited at the left hand end of the chain, say the site $j=0$, while its 
neighboring spin is $\vec{S}_1$. For $S=1/2$, the model is reduced to that considered in Ref.[21]. Interestingly when $S=1
/2$ or $S'=1/2$, the interaction term takes the simple form 
\begin{eqnarray}
H_{imp}=J\vec{S}_1\cdot\vec{S'},
\end{eqnarray}  
 with coupling constant $J={J_0}/[{(S'+S)^2-c^2}]$, which can  range from negative infinity to positive one, and meet 
 all the physical situations. So at least in these two cases, the Hamiltonian could be expected to properly describe the boundary bond effect in some real quasi-1D materials at very low temperature, such as the possible bond impurity $S'=1/2$ in $S=1$ Heisenberg antiferromagnet TMNIN\cite{27}.

To show the integrability of the Hamiltonian (7), let us first notice that the impurity term 
$H_{imp}$ could be more conveniently treated as the boundary operator, similar to the usual open
 boundary problem with an external field  applied to the end.  In addition to the Yang-Baxter equation (YBE) as the integrable
 condition of the bulk, there are some new consistent constraints (often 
 called the reflect YBE ) for the same model to be integrable under the open boundary conditions, and 
 the QISM is still available\cite{28}.  A new $K$-operator is introduced corresponding to the boundary 
 impurity. In most works, the $K$-operator is a $2\times 2$ matrix with the elements being $c$-number 
 describing the 
$S_z$-coupling to the applied field\cite{28}. The Sklyanin's formalism can be extended to the generic 
representations of $K$-operator, which is written as a $2\times 2$ matrix but with elements being 
operators rather than $c$-numbers\cite{20,21}. This operator-valued $K$ plays an useful role in 
constructing the boundary problem where the quantum degrees of freedom of the boundary enter 
interactions. Of course generally, both $R, K$-vertices could be interpreted as the 
inhomogeneous vertices in a 2D lattice model.
Our model corresponds to a very special one that the only ``inhomogeneity" comes from the 
 boundary row, leaving others uniform. It is built from 
\begin{eqnarray}
K(\lambda)=_{SS'}R^{A0}(\lambda-ic)_{SS'}R^{A0}(\lambda+ic)
\end{eqnarray}
and $_{SS'}R(\lambda)^{A0}$ is given by 
\begin{eqnarray}
_{SS'}R^{A0}(\lambda)=-\sum^{S+S'}_{l=|S-S'|}\prod^l_{k=|S-S'|+1}\frac{\lambda-k}{\lambda+k}
\prod^{S+S'}_{n\neq l, n=|S-S'|}\frac{y-y_n}{y_l-y_n}.
\end{eqnarray}
It is straightforward to show that the doubled monodromy matrix
\begin{eqnarray}
\Theta(\lambda)=T(\lambda)K(\lambda)T^{-1}(-
\lambda)
\end{eqnarray}
satisfies the reflection YB relations and its trace $\theta(\lambda)=tr_A\Theta(\lambda)$ satisfies   
$[\theta(\lambda), \theta(\mu)]=0, ~~\forall \lambda, \mu$.
Similarly, because
$H\propto \frac{d}{d\lambda}\ln \theta(\lambda)|_{\lambda=0}$,
the Hamiltonian (7) is indeed integrable. Its spectrum is uniquely determined by the following Bethe 
ansatz equations (BAE):
\begin{eqnarray}
\frac{\lambda_j+i(S'+c)}
{\lambda_j-i(S'+c)}\cdot
\frac{\lambda_j+i(S'-c)}
{\lambda_j-i(S'-c)}\cdot
[\frac{\lambda_j+iS}{\lambda_j-iS}]^{2N+1}
=\prod_{l\neq j}^M
\frac{\lambda_j-\lambda_l+i}{\lambda_j-\lambda_l-i}\cdot
\frac{\lambda_j+\lambda_l+i}{\lambda_j+\lambda_l-i}.
\end{eqnarray}
 The energy of the Hamiltonian (7) is 
\begin{eqnarray}
 E=-J_0\sum_{j=1}^M\frac{S}{\lambda_j^2+S^2}
\end{eqnarray}
up to a rapidity-independent constant. The magnetization is given by $S_z=NS+S'-M$ with $M$ being the 
number of down-spins.
It is recalled that when $S'=S$, $c=0$ (or $J=J_0$), the impurity is just the one more site of the 
chain. For convenience, we put $J_0=1$ in the following text. 
\section{Ground state, boundary correlator and boundary strings}
Due to the reflection symmetry of the model and its Bethe ansatz equation, there is a restriction on
the rapidities: $\lambda_j\neq\pm\lambda_l$, for $\j\neq l$. Therefore, $\lambda_j=0$ is forbidden
in this system. Generally, the bulk solutions of (13) can be described by the following strings 
in the thermodynamic limit
\begin{eqnarray}
\lambda_{j,\gamma}^n=\lambda_\gamma^n+\frac i2(n-2j+1), {~~~~}j=1,2,\cdots,n
\end{eqnarray}
with $\lambda_{\gamma}^n$ a positive real number. Since $c$ and $-c$ give the same Hamiltonian, not
lossing generality, we consider only $c>0$ ($c$ real) or $Im{~}c>0$ ($c$ imaginary) cases. For $c<S'$
and or imaginary $c$, (15) are the only possible solutions of the Bethe ansatz equation (13). For each class of states classified by
$n$-strings, we introduce the usual density distribution function $\rho_n(\lambda)$ and $\rho_{n,h}(\lambda)$,
representing occupied states (particles) and missing states (holes) respectively. The Bethe ansatz equation
of the $n$-strings reads:
\begin{eqnarray}
\rho_{n,h}(\lambda)+\sum_{l=1}^\infty{\bf A}_{nl}\rho_l(\lambda)=a_{n,2S}(\lambda)+\frac1{2N}[\phi_n^{imp}(\lambda)
+\phi_n^{edg}(\lambda)],
\end{eqnarray}
where $a_n(\lambda)=n/2\pi(\lambda^2+n^2/4)$, $a_{n,l}(\lambda)=\sum_{k=1}^{min(n,l)}a_{n+l+1-2k}(\lambda)$; ${\bf A}_{nl}$
is an integral operator with the kernel
$$
A_{n,l}(\lambda)=a_{|n+l|}(\lambda)+2\sum_{k=1}^{min(n,l)-1}a_{n+l-2k}(\lambda)+a_{|n-l|}(\lambda),
$$
$\phi_n^{imp}(\lambda)=a_{n,2S'}(\lambda-ic)+a_{n,2S'}(\lambda+ic)$ is the impurity contribution; 
$\phi_n^{edg}(\lambda)=a_{n}(\lambda)+a_{n+1}(\lambda)+a_{n-1}(\lambda)(1-\delta_{1,n})$
is the surface or edge term, which is  
independent of the magnetic impurity. Notice $\delta(\lambda)$ in $\phi^{edg}_n$ is included to cancel the
$\lambda_{\gamma}^n=0$ term, which is the solution of the Bethe ansatz equation but corresponds to a zero wave
function (a direct result of the restriction $\lambda_j\neq \lambda_l$). In the ground state, only $2S$-strings
exist\cite{26} and (16) is reduced to
\begin{eqnarray}
{\bf A}_{2S,2S}\rho_{2S}(\lambda)=a_{2S,2S}(\lambda)+\frac1{2N}[\phi_{2S}^{imp}(\lambda)
+\phi_{2S}^{edg}(\lambda)],
\end{eqnarray}
By Fourier transforming (17), we readily obtain
\begin{eqnarray}
\rho_{2S}(\lambda)=\rho_{2S}^0(\lambda)+\frac1{2N}[\rho_{2S}^{imp}(\lambda)+\rho_{2S}^{edg}(\lambda)],\\
\rho_{2S}^0(\lambda)=\frac 1{2\cosh(\pi\lambda)},\\
\rho_{2S}^{imp}(\lambda)=\frac1{2\pi}\int\frac{\sinh(S' \omega)\cosh(c\omega)}
{\cosh\frac\omega2\sinh(S\omega)}e^{-i\omega\lambda}d\omega,{~~~}for{~~} S>S'\\
\rho_{2S}^{imp}(\lambda)=\frac1{2\pi}\int\frac{e^{-(S'-S)|\omega|}\cosh(c\omega)}{\cosh\frac\omega 2}
e^{-i\omega\lambda}d\omega,{~~~}for{~~}S<S'\\
\rho_{2S}^{edg}(\lambda)=\frac1{2\pi}\int\frac{\tanh\frac\omega 2(1-e^{S|\omega|})}
{2\sinh(S\omega)}e^{-i\lambda\omega}d\omega+\rho_{2S}(\lambda).
\end{eqnarray}
The ground state energy can be readily calculated as
\begin{eqnarray}
E_0/N=f_{bulk}^0+\frac1N(F_{imp}^0+F_{edg}^0+\frac12 f_{bulk}^0),\\
f_{bulk}^0=\frac12[\Psi(\frac12)-\Psi(\frac12+S)],\\
F_{imp}^0=\frac14\sum_{r=\pm}[\Psi(\frac12+\frac12|S-S'|+irc)-\Psi(\frac12+\frac12(S+S')+irc)],\\
F_{edg}^0=\frac14[\Psi(\frac12+\frac12(S-\frac12))-\Psi(\frac12+\frac12(S+\frac12))\nonumber\\
+\Psi(\frac12+\frac12|S-1|)-\Psi(\frac12+\frac12(S+1))].
\end{eqnarray}
where $\Psi$ is the digamma function. As a byproduct, the correlator of ${\vec S}'$ and ${\vec S}_1$
can be exactly derived for the present model. When $S=1/2$ or $S'=1/2$ we have
\begin{eqnarray}
<{\vec S}'\cdot{\vec S}_1>=\frac {\partial}{\partial J} E_0.
\end{eqnarray}
Since $c$ (and therefore $J$) is only included in $\rho_{2S}^{imp}(\lambda)$,
we need only the impurity energy $F_{imp}^0$. The boundary correlator can be calculated  as
\begin{eqnarray}
<{\vec S}'\cdot{\vec S}_1>=J^{-2}\frac{\partial}{\partial c^2}F_{imp}^0.
\end{eqnarray}

Now we turn to $c>S'$ case. In addition to the $n$-string solutions (15), an imaginary mode $\lambda=i(c-S')$ appears to be
a solution of the Bethe ansatz
equation (13). In fact, in this case a so-called boundary $n-k$-string\cite{29} is possible solution of the BAE
\begin{eqnarray}
\lambda_{bs}^{n-k,m}=i(c-S')+im,{~~~~~~~}m=k,k+1,\cdots,n,
\end{eqnarray}
where $k<S'-c$ or $k=0$. Generally, there is no restriction for $n$ in the spin chain with a boundary field\cite{29}. However,
in our case, $\lambda=\pm i(S'+c)$ are not solutions as we can see from the Bethe ansatz equation. That means $n<2S'$. In
addition, for $2c=integer$ case, $k$ must be zero due to the restriction $\lambda_j\neq \lambda_l, j\neq l$. Suppose there is
an $n-k$ boundary string in the ground state configuration. The change of the $2S$-string distribution $\rho_{bs}(\lambda)/2N$
can be readily derived from the following equation
\begin{eqnarray}
{\bf A}_{2S,2S}\rho_{bs}(\lambda)=-\sum_{l=k}^n\{a_{2S,2S}[\lambda-i(c-S'+l)]+a_{2S,2S}[\lambda+i(c-S'+l)]\}.
\end{eqnarray}
The energy carried by the boundary string is
\begin{eqnarray}
\epsilon_{bs}=-\frac\pi 2\int a_{2S,2S}(\lambda)\rho_{bs}(\lambda)-\sum_{l=k}^n\frac S{S^2-(c-S'+l)^2}.
\end{eqnarray}
By solving (30) via Fourier transformation and submitting $\rho_{bs}$ into (31), we find $\epsilon_{bs}=0$. That means
the boundary string contributes nothing to the total energy in the thermodynamic limit. It corresponds to a charged
vacuum in the sine-Gordon theory\cite{29}. We remark that some boundary string will be stablized with a finite magnetization.
Such a situation is much like those of the fermion systems with boundary potential\cite{29} or Kondo impurity\cite{20,22}.
\section{Thermodynamics} 
In this section, we consider the thermodynamics of $c\leq S'$ (antiferromagnetic) case. The thermodynamic Bethe ansatz equations can be derived by following the standard method\cite{8,30,31}. At finite temperatures,
the solutions of BAE are described by (15) . The energy of the system takes the form
\begin{eqnarray}
\frac E N=-\pi\sum_{n=1}^\infty\int a_{n,2S}(\lambda)\rho_n(\lambda)d\lambda+\sum_{n=1}^\infty nH\int \rho_n(\lambda)d\lambda,
\end{eqnarray}
where $H$ is the external magnetic field. The entropy of the system reads
\begin{eqnarray}
S/N=\sum_{n=1}^\infty\int\{(\rho_n+\rho_{n,h})\ln(\rho_n+\rho_{n,h})-\rho_n\ln\rho_n-\rho_{n,h}\ln\rho_{n,h}\}d\lambda.
\end{eqnarray}
By minimizing the free energy $F=E-TS$ we readily obtain the following equation
\begin{eqnarray}
\ln(1+\eta_n)=\frac{nH-\pi a_{n,2S}}T+\sum_{l=1}^\infty {\bf A}_{n,l}\ln(1+\eta_l^{-1}),
\end{eqnarray}
where $\eta_n(\lambda)\equiv \rho_{n,h}(\lambda)/\rho_n(\lambda)$ and $\eta_0(\lambda)\equiv 0$. With the identities
\begin{eqnarray}
{\bf A}_{n,m}-{\bf G}({\bf A}_{n-1,m}+{\bf A}_{n+1,m})=\delta_{n,m}, {~~~~~}{\bf A}_{1,m}-{\bf G}{\bf A}_{2,m}=\delta_{1,m},\\
{\bf B}_{n,m}-{\bf G}({\bf B}_{n-1,m}+{\bf B}_{n+1,m})=\delta_{n,m}{\bf G}, {~~~~~}{\bf B}_{1,m}-{\bf G}{\bf B}_{2,m}=\delta_{1,m}
{\bf G},
\end{eqnarray}
where ${\bf B}_{n,m}$ and ${\bf G}$ are integral operators with the kernels $a_{n,m}(\lambda)$ and $1/2\cosh(\pi \lambda)$
respectively, (34) can be reduced to
\begin{eqnarray}
\ln\eta_n=-\frac \pi{2T\cosh(\pi\lambda)}\delta_{n,2S}+{\bf G}[\ln(1+\eta_{n-1})+\ln(1+\eta_{n+1})],
\end{eqnarray}
with the boundary condition
\begin{eqnarray}
\lim_{n\to\infty}\frac{\ln\eta_n}n=\frac HT.
\end{eqnarray}
(37) is almost the same equation to that of the $2S$-channel Kondo problem\cite{8} with only a different driving term.
The free energy reads
\begin{eqnarray}
F/N=f_{bulk}+\frac 1NF_{imp}+\frac1NF_{edg},\\
f_{bulk}=f_{bulk}^0-T\int[2\cosh(\pi\lambda)]^{-1}\ln[1+\eta_{2S}(\lambda)]d\lambda,\\
F_{edg}=\frac14H-\frac12T\int[2\cosh(\pi\lambda)]^{-1}\{\ln[1+\eta_1(\lambda)]+
\ln[1+\eta_2(\lambda)]\}d\lambda,\\
F_{imp}=-(S'-\frac12)H-\frac12T\sum_{n=1}^\infty\int\Phi_{n}^{imp}(\lambda)\ln[1+\eta_n^{-1}(\lambda)]d\lambda,
\end{eqnarray}
It is not easy to reduce further $F_{imp}$ for arbitrary real $c$. The boundary behaves always as a spin-1/4 (in fact one half
of a spin-1/2) and its critical effect in the $XXZ$ spin-1/2 chain has been discussed in a previous work\cite{32}. In our case,
the ``boundary spin" shows overscreened critical behavior as will be discussed in the following text. 
\subsection{High Temperature Limit}
When $T\to\infty$, the driving term tends to zero. Therefore in this limit the functions $\eta_n(\lambda)$ 
tend to constants $\eta_n^{-}$  satisfing
 the algebraic equation\cite{31,8}
\begin{eqnarray}
{\eta_n^-}^2=(1+\eta_{n-1}^-)(1+\eta_{n+1}^-),
\end{eqnarray}
with the boundary conditions
\begin{eqnarray}
\eta_0^-\equiv0,{~~~~~}\lim_{n\to\infty}\frac{\ln\eta_n^-}n=\frac HT\equiv 2x_0.
\end{eqnarray}
The solutions of (43) are
\begin{eqnarray}
\eta_n^-=\frac{\sinh^2(n+1)x_0}{\sinh^2x_0}-1.
\end{eqnarray}
Substituting (45) into (41) we get
\begin{eqnarray}
F_{edg}\sim-\frac12T\ln(2\cosh x_0).
\end{eqnarray}
Notice that only the pure boundary term is given in the above expression. Obviously, the open boundary's contribution to the free energy is one half of a spin-1/2. The entropy of the open boundary is
$\ln\sqrt{2}$. To calculate the impurity free energy, we note that the integral kernel $\Phi_{n}^{imp}$ in (42) can be 
written with real variable as
\begin{eqnarray}
\Phi_{n}^{imp}(\lambda)=\sum_{r=\pm}\sum_{k=1}^na_{|n+1+2S'+2rc-2k|}(\lambda)\epsilon(n+1+2S'+2rc-2k),
\end{eqnarray}
where $\epsilon(x)=sign(x)$ and $\epsilon(0)=0$. Since all $a_n(\lambda)$ are equivalent to $\delta(\lambda)$ in $T\to\infty$
limit, we can replace (47) by
\begin{eqnarray}
\Phi_{n}^{imp}(\lambda)\to a_{n,2S'-c_I}(\lambda)+
a_{n,2S'+c_I}(\lambda),{~~~~~~}for {~~~}c_I=2c\\
\Phi_{n}^{imp}(\lambda)\to a_{n,2S'-c_I}(\lambda)+
a_{n,2S'+c_I}(\lambda)+\delta(\lambda)\sum_{l=1}^{c_I}\delta_{n,2S'-c_I+2l-1},{~~~~~}for{~~~}c_I\neq 2c,
\end{eqnarray}
where $c_I$ is the integer part of $2c$. Therefore, the impurity free energy reads
\begin{eqnarray}
F_{imp}\sim-\frac14T[\ln(1+\eta_{2S'-c_I}^-)+
\ln(1+\eta_{2S'+c_I}^-)], {~~~~}for{~~~~}c_I=2c,\\
F_{imp}\sim-\frac14T[\ln(1+\eta_{2S'-c_I}^-)+
\ln(1+\eta_{2S'+c_I}^-)]\nonumber\\
+\frac12T\sum_{l=1}^{c_I}[\ln(1+\eta_{2S'-c_I+2l-1}^-)-\ln\eta_{2S'-c_I+2l-1}^-],{~~~~}for {~~~}c_I\neq 2c.
\end{eqnarray}
 The entropy of the impurity is
\begin{eqnarray}
S_{imp}=\ln\sqrt{(2S'+1)^2-c_I^2}, {~~~~}for{~~~}c_I=2c,\\
S_{imp}=\ln\sqrt{(2S'+1)^2-c_I^2}\nonumber\\
-\sum_{l=1}^{c_I}\ln\frac{2S'-c_I+2l}{\sqrt{(2S'-c_I+2l-1)(2S'-c_I+2l+1)}}, {~~~}for{~~~}c_I\neq 2c.
\end{eqnarray}
>From (52-53) we see that even at high temperature limit, the impurity spin is split into ghost spins via the boundary 
bond deformation. Though the magnetization of the impurity is almost the same to that of a free spin ${\vec S}'$, the entropy
is strongly interaction-dependent.
\subsection{Low Temperature Limit}
When $T\to 0$, the driving term in (37) diverges. That means $\eta_{2S}\to 0$ and all other $\eta_n$ tend to constant $\eta_n^+$
which satisfy the same equation of $\eta_n^-$ but with a different boundary condition. Since the equation is decoupled at $n=2S$, we
have different solutions\cite{8} for $n>2S$ and $n\leq 2S$
\begin{eqnarray}
\eta_n^+=\frac{\sinh^2(n-2S+1)x_0}{\sinh^2x_0}-1, {~~~~~} for{~~~}n\geq 2S,\\
\eta_n^+=\frac{\sin^2\frac\pi2\frac{n+1}{S+1}}{\sin^2\frac\pi{2(S+1)}}-1,{~~~~~}for{~~~}n<2S.
\end{eqnarray}
The residual entropy of the open boundary is
\begin{eqnarray}
S_{edg}=\frac12\ln[2\cos\frac\pi{2(S+1)}].
\end{eqnarray}
For $2c=c_I$, the impurity behaves still as two ghost spins $S'+c_I/2$ and $S'-c_I/2$. The entropy of a ghost spin ${\bar S}$ reads
\begin{eqnarray}
S_{ghost}=\frac12\ln[2({\bar S}-S)+1], {~~~~} for{~~~} |{\bar S}|>S,\\
S_{ghost}=\frac12\ln\frac{\sin\frac\pi2\frac{2{\bar S}+1}{S+1}}{\sin\frac\pi{2(S+1)}},{~~~~}for{~~~}
|{\bar S}|\leq S.
\end{eqnarray}
The summation of the two ghost spins' entropy gives that of the whole impurity. When $c_I\neq 2c$, the difference between the 
residual entropies for $c_I\neq 2c$ and $c_I=2c$, i.e., $\Delta S_{imp}$ reads
\begin{eqnarray}
\Delta S_{imp}=-\frac12\sum_{l=1}^{c_I}[\ln(1+f_{2S'-c_I+2l-1})-\ln f_{2S'-c_I+2l-1}],
\end{eqnarray}
with $f_n=\lim_{x_0\to 0}\eta_n^+$. This result shows that the spin configuration of the ground state is very complicated
and strongly depends on the impurity-bulk coupling. It would be plausible to coin it as a local spin glass. In fact, the residual
entropy has  jumps at $c=c_I/2$. That means quantum phase transition occurs for $c$ across a half integer or an
integer.\par 
To obtain the leading order of some thermodynamic quantities such as the specific heat and the susceptibility, we need the
low temperature ($T<<T_k$) expansion. This can be done by following the standard method developed for the multi-channel
Kondo problem\cite{8}. 
For $T\to 0$, only the excitations near the Fermi surface ($\lambda\to \pm\infty$) are important. The driving term in (37) 
can be approximately replaced by $-(\pi/T)exp(-\pi|\lambda|)$. We introduce the new variables $\zeta_\pm=\pm\pi\lambda+\ln(\pi/T)$,
then $\eta_n$ take the following asymptotic form\cite{8}
\begin{eqnarray}
\eta_n(\zeta_\pm)\sim\eta_n^++(\alpha_n+\beta_nx_0^2)e^{-\zeta_\pm},{~~~} for{~~~}n\geq 2S,\\
\eta_n(\zeta_\pm)\sim\eta_n^++(\alpha_n+\beta_nx_0^2)e^{-\tau\zeta_\pm},{~~~} for{~~~}n< 2S.
\end{eqnarray}
Here $\alpha_n$ and $\beta_n$ are constants, $\tau=2/(S+1)$ and $\pm$ denotes the two Fermi points. For imaginary $c=ib$, the 
free energy of the impurity reads
\begin{eqnarray}
F_{imp}\sim F_{imp}^0-T\int[\frac 1{2\cosh(\zeta+\pi b-\ln\frac\pi T)}+\frac 1{2\cosh(\zeta-\pi b-\ln\frac\pi T)}]
\ln(1+\eta_{2S'})d\zeta.
\end{eqnarray}
Notice that we have replaced $\zeta_\pm$ by $\zeta$ in the integral. In this case, the bond deformation does not change the effective strength
of the impurity but the energy scale $T_k$ (Kondo temperature)\cite{21}
\begin{eqnarray}
T_k\sim\pi\cosh^{-1}(\pi b).
\end{eqnarray}
The system behaves as a $2S$-channel Kondo system with an impurity $\vec{S}'$. For real $c$ and $c_I=2c$, the free energy of the impurity
can be rewritten as
\begin{eqnarray}
F_{imp}\sim F_{imp}^0-T\sum_{\pm}\int \frac{\ln(1+\eta_{2S'\pm c_I})} {2\cosh(\zeta-\ln\frac\pi T)}d\zeta.
\end{eqnarray}
For $c_I\neq 2c$, 
\begin{eqnarray}
F_{imp}\sim F_{imp}^0-T\sum_{\pm}\int G_\pm[\frac 1\pi(\zeta-\ln\frac\pi T)]\ln(1+\eta_{2S'\pm c_I})d\zeta\nonumber\\
-\frac T{2\pi}\sum_{l=1}^{c_I}\int\frac{c-\frac12c_I}{[\frac1\pi(\zeta-\ln\frac \pi T)]^2+(c-\frac12 c_I)^2}\ln(1+\eta_{2S'-c_I
+2l-1}^{-1})d\zeta,
\end{eqnarray}
where
\begin{eqnarray}
G_\pm(\lambda)=\int\frac{e^{\mp(c-\frac12 c_I)|\omega|}}{4\pi\cosh\frac \omega 2}e^{-i\lambda\omega}d\omega.
\end{eqnarray}
Notice $G_\pm(\lambda)$ are convergent in the real axis since $c-c_I/2<1/2$. The specific heat and the susceptibility can be easily
derived from the free energy. With different values of $c$, different quantum critical behavior may appear. (i) For
$S\pm c_I/2\geq S$, both the ghost spins are underscreened and the leading terms in the specific heat and the susceptibility
are the Schottky term and the Curie term respectively. (ii) For $S'-c_I/2<S<S'+c_I/2$, no matter how large $S'$ is,
\begin{eqnarray}
C_{imp}\sim T^\tau,{~~~~~~}\chi_{imp}\sim T^{-1}+O(T^{\tau-1}),
\end{eqnarray}
which indicate a novel critical behavior. (iii) For $S'\pm c_I/2<S$, the system behaves as a conventional overscreened Kondo 
system. The above results show that the bond deformation has two effects: The half integer part of $c$ ($c_I/2$) renormalizes the 
impurity spin and the rest ($c-c_I/2$) renormalizes the effective energy scale $T_k$ (Kondo temperature).
\section{Concluding remarks}
In conclusion, we propose an integrable model of a boundary impurity spin ${\vec S}'$ coupled with an open Takhatajian-Babujian
spin-$S$ chain. The relation between the present model and the bulk impurity problem in a spin chain is discussed. In our
model, when $S$ or $S'$ is one half,  The ``fine tuned" effect in the periodic integrable models\cite{5,6} is overcome and the interaction term in our
case takes a very simple form. The coupling constant $J$ can take arbitrary value without destroying the integrability of
the Hamiltonian, while in the periodic models, there is a constraint to $J$. Though a similar $_{S'S}R^{A0}(\lambda-c)$
can be introduced in the periodic models\cite{6,5}, the parameter $c$ must be real (imaginary in our case) which describes a
weak-linked impurity to the bulk. The interaction only affects the energy scale (Kondo temperature) but does not change
the fixed point of the system. With an imaginary $c$, the model Hamiltonians constructed for bulk impurities are non-Hermitian
and their spectra generally lie in the complex plane\cite{33} rather than in the real axis completely. In our model, both
real $c$ and imaginary $c$ define a Hermitian Hamiltonian due to the reflection symmetry, and the coupling constant $J$ meets
all physical situations. Some new quantum critical phenomena driven by the impurity-bulk coupling have been found, which can
never appear in the periodic models:(i)The stronger coupling $J$ may split the impurity spin into effective ``ghost spins"
$S'-c_I/2$ and $S'+c_I/2$. That means the coupling not only change the energy scales (Kondo temperature) as in the conventional
Kondo problem but also renormalizes the effective strength of the impurity spin. It seems that the strength of these
ghost spins does not change via temperature. Such a phenomenon reveals a pure correlation effect. We remark a similar effect can
be induced by the impurity potential in the Luttinger-Kondo problems\cite{20,22}.(ii)Depending on the strength of the coupling,
the system may show a variety of critical behavior differing from those of the conventional Kondo problems. A typical example
is that when $S'-c_I/2<S<S'+c_I/2$, the leading term in the susceptibility is Curie type, while that of the specific heat is 
overscreened $2S$-channel Kondo type. Such a fascinating non-Fermi liquid behavior has never been found in the conventional
impurity problem (notice these are induced by the same impurity). (iii)The open boundary, which can be produced by either an
impurity (no matter magnetic or non-magnetic) or bond deformation, shows overscreened multi-channel behavior
as long as the bulk spin $S>1/2$. Such an effect is caused by the self-avoiding of the scattering of each spin-wave with its 
reflection counterpart and is common to the multi-channel systems in 1D. Our results strongly suggests that some new
intermediate fixed point may exist for the Kondo problem in a strongly correlated host.
\par
Two of the authors (JD, YW) are grateful to Profs. N. Andrei and Yu Lu for helpful discussions. YW also acknowledges the financial support of the Alexander von Humboldt-Stiftung and China National Foundation for Natural Science.

\end{document}